\documentclass[aps,showpacs,twocolumn]{revtex4}
\usepackage{epsfig}
\usepackage{color}
\usepackage{amsmath}
\usepackage{amssymb}
\usepackage{bbm}

\newcommand{\be}{\begin{equation}}
\newcommand{\ee}{\end{equation}}
\newcommand{\bea}{\begin{eqnarray}}
\newcommand{\beaa}{\begin{eqnarray*}}
\newcommand{\eea}{\end{eqnarray}}
\newcommand{\eeaa}{\end{eqnarray*}}

\usepackage{graphicx} 

\begin{document}

\title{The impact of the electric field on superconductivity in the time-dependent Ginsburg-Landau theory.}

\author{N. Karchev and T. Vetsov }

\affiliation{Department of Physics, University of Sofia, 1164 Sofia, Bulgaria}


\begin{abstract}

In this letter, we address the impact of the electric field on superconductors who are insulators in the normal state, semiconductors at low carrier concentration and ultracold gas of fermions in the strongly interacting regime.  The electric field penetrates these systems and effects on the Cooper pairs. We show that if there are Cooper pairs above the superconductor critical temperature the electric field forces the Cooper pairs to Bose condensate and the onset of the superconductivity, thereby increasing the critical temperature. To study this phenomenon we numerically solve the Maxwell equations for s-wave superconductors obtained from the time-dependent Ginsburg-Landau theory. Our investigation designs an experimental way for verification of the pairing of Fermions preceding superconductivity and superfluidity.

\end{abstract}


\maketitle

The physical origin of the superconducting phenomenon is the attractive interactions between fermions. They form "bosonic-like" Cooper pairs which are driven to Bose condensate. In the J. Bardeen, L. N. Cooper and J. R. Schrieffer (BCS) theory \cite{BCS57} the condensation and pair formation temperatures coincide. In a paper, \cite{Eagles69} Eagles studied the possibility of pairing without superconductivity in some superconducting semiconductors at low carrier concentration. Later on A. J. Leggett \cite{Laggett80} addressed the problem in a dilute gas of fermions at zero temperature motivated by superfluid $^3$He. P. Nozi\`{e}res and S. Schmitt-Rink extended his analysis to other models and to finite temperature \cite{NS-R85}.

The above materials have a low concentration of charge carriers and a strong interaction between them. As a result pairing of fermions prior to superconductivity is possible.
This scenario fits well for the description of the underdoped phase of high-temperature cuprates superconductors. The anomalously short coherence length combined with the observation of pseudo-gap has motivated a number of authors \cite{Ullah91, Chen98, Larkin02, Iddo02} to expect stronger than BCS attraction so that the bosonic degrees of freedom appear at a temperature T* which may be significantly larger than $T_c$.
The fluctuations of Cooper pairs above $T_c$ lead to charge transfer. This so-called paraconductivity is discussed in \cite{Ullah91, Chen98, Larkin02}. The fluctuations can be responsible for increasing, in the vicinity of the transition temperature, of the heat capacity, the diamagnetic susceptibility, etc\cite{Larkin02}. Their contribution to the thermal transport in the normal state is theoretically analyzed in \cite{Iddo02}.

The experiments with underdoped copper oxide $Bi_2Sr_2CaCu_2O_8$ \cite{Yang08,Kondo11}
show that the pseudogap does reflect the formation of preformed pairs of electrons.

An important example is the sulfur hydride. It is a superconductor at very high pressure and temperature \cite{Drozdov15}. The temperature dependence of the resistance, measured at different pressures, shows that the material undergoes a superconductor-insulator transition below 129 GPa.

Atomic quantum gases have been proposed as quantum simulators to identify the microscopic origin of condensed
matter phenomena, which have been pondered for decades.
Experiments with cold atomic gases provide a remarkable alternative to tackle the question of interactions in quantum fermi systems.
In cold-atom experiments,
there is a powerful tool to control the two-body coupling strength and sign in the system  \cite{Inouye98}.

The developments in experiments with low density ultracold degenerate Fermi gases led to the observation of pairing of fermions \cite{Greiner04, Zwierlein04} without superfluidity.
The measurement of the momentum distribution of fermionic atom pairs, as a function of temperature, in \cite{Regal04} demonstrated the Bose condensation of pairs and superfluidity in the system. An alternative method is used in \cite{Zwierlein05}.

In the present paper, we study the effect of the applied electric field on the fermion pairs prior to the superconductivity. To this end, we consider the time-dependent Ginsburg-Landau (TDGL) theory.  The theory is widely used to study the fluctuations of Cooper pairs above the superconducting critical temperature \cite{Ullah91, Chen98, Larkin02, Iddo02, Melo93, Shina04, Aoki15}.  We investigate a model with field-theory action \cite{Gorkov68, Thompson70, Tinkham75}
\bea\label{MSc14}
S & = & \int d^4x\left [-\frac 14 \left (\partial_{\lambda}A_{\nu}-\partial_{\nu}A_{\lambda}\right)\left (\partial^{\lambda}A^{\nu}-\partial^{\nu}A^{\lambda}\right)\right. \nonumber\\
& + & \left.  \frac {1}{D}\psi^*\left (i\partial_{t}-e^*\varphi\right)\psi \right. \\
& - & \left.  \frac {1}{2m^*} \left (\partial_{k}-ie^*A_{k}\right)\psi^* \left (\partial_{k}+ie^*A_{k}\right)\psi \right. \nonumber \\
& + & \left. \alpha \psi^*\psi -\frac {g}{2}\left(\psi^*\psi \right)^2\ \nonumber \right],
\eea
written in terms of gauge four-vector electromagnetic potential $"A"$ and complex scalar field $"\psi"$-the superconducting order parameter. We use the notations $x=(x^0,x^1,x^2,x^3)=(\upsilon t,x,y,z)$, $\upsilon^{-2}=\mu\varepsilon$, where $\mu$ is the magnetic permeability and $\varepsilon$ is the electric permittivity of the superconductor.  The constant $D$ is the normal-state diffusion, ($e^*,m^*$) are effective charge and mass of superconducting quasi-particles and $\varphi=\upsilon A_0$ is the electric scalar potential. The index $k$ runs $k=x,y,z$ and $\alpha$ is a function of the temperature $T$
\begin{equation}
\alpha=\alpha_0(T_c-T),\end{equation}
where $T_c$ is the superconducting critical temperature and $\alpha_0$ is a positive constant.

The action (\ref{MSc14}) is invariant under the gauge transformations
\begin{eqnarray}
\psi'(x) & = & \exp {[i e^*\phi(x)]}\psi \nonumber \\
A'_{\nu} & = & A_{\nu}-\partial_{\nu}\phi(x),
\end{eqnarray}
where $\phi(x)$ is a real function. We represent $\psi(x)$ in the form
$\psi(x) =  \rho(x)\exp {[ie^*\theta(x)]}$,  where $\rho(x)=|\psi(x)|$ is a gauge invariant and the gauge transformation of $\theta(x)$ is
$\theta'(x) =  \theta(x)+\phi(x)$. Thereafter, one rewrites the equation (1) in terms of $\rho(x)$ and $\theta(x)$.

It is convenient to use the action in the first-order formalism \cite{SF} as gauge potential $A^{\lambda}$, phase $\theta$,  gauge invariant antisymmetric field $F^{\lambda\nu}=-F^{\nu\lambda}$, gauge invariant vector field $Q_k$ and gauge invariant scalar field $\rho$ are considered to be independent degrees of freedom in the theory.
\bea\label{MSc15}
S & = & \int d^4x\left \{-\frac 12 \left [\left (\partial_{\lambda}A_{\nu}-\partial_{\nu}A_{\lambda}\right)F^{\lambda\nu}
-\frac 12 F_{\lambda\nu}F^{\lambda\nu}\right]\right. \nonumber \\
& - & \left. \frac {e^*}{D}\rho^2\left (\varphi+\partial_t\theta\right)\right. \\
& - & \left.  \frac {e^{*2}}{m^*}\rho^2\left[\left (\partial_{k}\theta+ A_{k}\right)Q_{k}-\frac 12 Q_{k}Q_{k}\right] \right. \nonumber  \\
& - & \left. \frac {1}{2m^*}\partial_{k}\rho\partial_{k}\rho +  \alpha \rho^2 -\frac {g}{2}\rho^4  \right\}.\nonumber
\eea
Varying the action with respect to the independent fields one obtains the system of equations:
\bea
& & F_{\lambda\nu}\, = \, \left (\partial_{\lambda}A_{\nu}-\partial_{\nu}A_{\lambda}\right)\label{MSc161} \\
& & Q_{k}\, = \,\partial_{k}\theta+ A_{k}\label{MSc162} \\
& & \partial_{\lambda}F^{\lambda}_{\,\,\,\,\,k}\, + \, \frac {e^{*2}}{m^*}\rho^2Q_{k} = 0 \label{MSc163} \\
& & \partial_{k}F_{0k}\, - \, \frac {\upsilon e^{*}}{D}\rho^2 = 0 \label{MSc164} \\
& & \partial_{t}\rho^2+\frac {D}{m^*}\partial_{k}\left(\rho^2Q_{k}\right) \, = \, 0 \label{MSc165}\\
& & \frac{1}{2m*}\Delta\rho+\alpha \rho-g\rho^3 -\frac {e^*}{D}\rho\left(\varphi+\partial_{t}\theta\right)                                      \nonumber \\
& & = \frac {e^{*2}}{m^*}\rho\left[\left (\partial_{k}\theta+ A_{k}\right)Q_{k}-\frac 12 Q_{k}Q_{k}\right].\label{MSc166}
\eea

If we substitute in equations (\ref{MSc163},\ref{MSc164},\ref{MSc165},\ref{MSc166}), the expressions for $F_{\lambda\nu}$ from (Eq.\ref{MSc161}) and $Q_k$ from (Eq.\ref{MSc162}) we obtain the system of equations following from the action (\ref{MSc14}) . This means that theories with the actions (\ref{MSc14}) and (\ref{MSc15}) are equivalent.

Alternatively, we eliminate the gauge fields $A_{\lambda}$ and $\theta$ from the system of equations and define a new gauge invariant field $Q$
\be\label{MSc17}
Q\, = \,\partial_{t}\theta+ \varphi,\ee
to obtain the system of equations for the gauge-invariant fields.
We use the standard representation for the tensor $F_{\lambda\nu}$ by means of the electric $\bf E$ and magnetic $\bf B$ fields:
$(F_{01},F_{02},F_{03})=\textbf{E}/\upsilon$, $(F_{32},F_{13},F_{21})=\textbf{B}$ and $(Q^0,Q^1,Q^2,Q^3)=(Q/\upsilon,\textbf{Q})$. In terms of
$\textbf{E},\textbf{B},\textbf{Q}$, $Q$ and $\rho$ the system of equations which describes the electrodynamics of s-wave superconductors is:
\begin{eqnarray}
& & \overrightarrow{\nabla}\times\textbf{B}\,=\,\mu\varepsilon\frac {\partial \textbf{E}}{\partial t}-\frac {e^{*2}}{m^*}\rho^2\textbf{Q} \label{EScCp11}\\
& & \overrightarrow{\nabla}\times\textbf{Q}\,=\,\textbf{B}\label{EScCp12}\\
& & \overrightarrow{\nabla}\cdot\textbf{E}\,=\,\frac {\mu\varepsilon e^{*}}{D}\rho^2 \label{EScCp13}\\
& & \overrightarrow{\nabla} Q+\frac {\partial \textbf{Q}}{\partial t}\,=\,-\textbf{E}\label{EScCp14}\\
& & \frac {1}{2m^*}\Delta\rho +\alpha \rho-g\rho^3- \frac {e^{*}}{D} \rho Q-\frac {e^{*2}}{2m^*}\rho \textbf{Q}^2=0.\label{EScCp15}
\end{eqnarray}

It is important to stress that the gauge-invariant vector $\textbf{Q}$ and scalar $Q$  take part in the equations (\ref{EScCp12}) and (\ref{EScCp14}) as a magnetic vector and electric scalar potentials, while in equation (\ref{EScCp11}) $(-2e^{*2}\rho^2\textbf{Q})$ is a supercurrent. This dual contribution of the new fields is the basis of the electrodynamics of superconductors.

We are interested in the system of equations for time-independent fields without magnetic field $\textbf {B}=\textbf{Q}=0$
\begin{eqnarray}
& & \overrightarrow{\nabla}\cdot\textbf{E}\,=\,\frac {\mu\varepsilon e^{*}}{D}\rho^2, \label{EScCp20}\\
& & \overrightarrow{\nabla} Q\,=\,-\textbf{E}\label{EScCp21}\\
& & \frac {1}{2m^*}\Delta\rho +\alpha \rho-g\rho^3- \frac {e^{*}}{D} \rho Q=0. \label{EScCp22}
\end{eqnarray}

\textcolor[rgb]{1.00,0.00,0.00}{To gain insight into the impact of the electric field on the superconductivity we do a qualitative analysis of the equation (\ref{EScCp22}) replacing the scalar field $Q$ by its average value $<Q>$. The electric scalar potential effectively changes the $\alpha$ parameter
$\alpha\rightarrow\alpha-\frac {e^{*}}{D}<Q>=\alpha_r$. If $<Q>$ is positive, the applied electric field decreases the $\alpha$ parameter and destroys superconductivity, while for negative $<Q>$ the parameter $\alpha$ increases. If the system is in a normal state $T>T_c$ and parameter $\alpha$ is negative, one can apply an electric field, strong enough, to change the sign of the renormalized parameter $\alpha_r>0$ which leads to Bose-condensation and onset of superconductivity.}

We consider a superconductor with slab geometry. The fields depend on one of the coordinates "z"  and the electric field vector has one nonzero component $\textbf{E}=(0,0,E)$. For a system in normal state, with negative $\alpha$, we arrive at the system of equations
\begin{eqnarray}
& & \frac{dE}{dz}\,=\frac {\mu\varepsilon e^{*}}{D}\rho^2, \nonumber \\
& & \frac{dQ}{dz}\,=\,-E \label{ESc60} \\
& & \frac {1}{2m^*}\frac {d^2\rho}{d^2z} -|\alpha| \rho-g\rho^3- \frac {e^{*}}{D} \rho Q=0. \nonumber
\end{eqnarray}

It is convenient to introduce dimensionless functions $f_1(\zeta),f_2(\zeta)$ and $f_3(\zeta)$ of a dimensionless distance $\zeta=z/\xi_{GL}$, where
\begin{equation}\label{MScApp9}\xi_{GL}=1/\sqrt{2m^*|\alpha|}\end{equation} and
\begin{eqnarray}\label{MScApp9}
Q(\zeta) & = & -E_0 \xi_{GL}f_1(\zeta), \nonumber \\
E(\zeta) & = &  E_0 f_2(\zeta),  \\
\rho(\zeta) & = & \rho_0 f_3(\zeta). \nonumber  \end{eqnarray}
In equations (\ref{MScApp9}) $\rho_0=\sqrt{|\alpha|/g}$ and the applied electric field is $\textbf{E}_0=(0,0,E_0)$. Finally, we introduce fourth function $f_4(\zeta)= df_3(\zeta)/d\zeta$. In terms of the functions $f_1,f_2,f_3,f_4$ the system of equations (\ref{ESc60}) adopts the form:
\begin{eqnarray}
& & \frac{df_1}{d\zeta}\,=\,f_2 \label{ESc51a}  \\
& & \frac{df_2}{d\zeta}\,=\frac {\kappa}{\gamma}f_3^2, \label{ESc51b}  \\
& & \frac{df_3}{d\zeta}\,=\, f_4 \label{ESc51c} \\
& & \frac {df_4}{d\zeta}\,-\,f_3\,-\,gf_3^3\,+\,\gamma f_1 f_3 =0,\label{ESc51d}
\end{eqnarray}
where
\begin{eqnarray}
& & \gamma\,=\,\frac {e^*\xi_{GL}}{D|\alpha|} E_0 \label{ESc52}\\
& & \kappa\,=\, \frac {e^{*2}\xi_{GL}^2\rho_0^2}{D^2\upsilon^{2}|\alpha|}\label{ESc53}
\end{eqnarray}

\begin{figure}[!ht]
\epsfxsize=\linewidth
\epsfbox{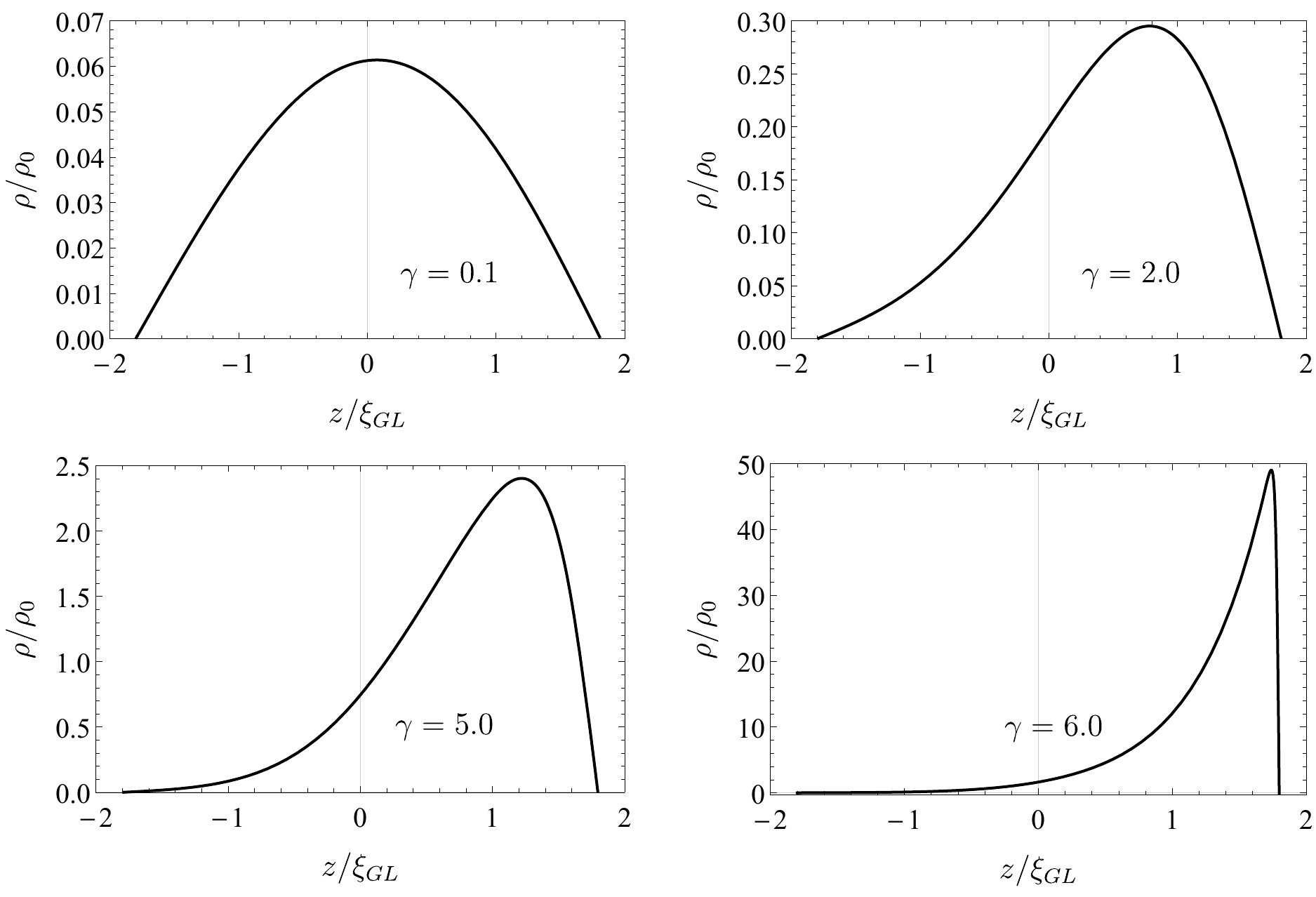} \caption{\,\,The dimensionless function $\rho/\rho_0$ of a dimensionless distance $\zeta=z/\xi_{GL}$, where $\rho_0=\sqrt{|\alpha|/g}$ and $\xi_{GL}=1/\sqrt{2m^*|\alpha|}$, is plotted. Increasing the applied electric field, keeping the parameter $\kappa$ fixed, $\kappa=2.5$, one increases the parameter $\gamma$. Figures show that applying electric field the Cooper pairs Bose condense $\rho/\rho_0>0$ and the onset of superconductivity starts while the temperature is higher then critical one $\alpha<0$. Increasing $\gamma$ increases the density $\rho/\rho_0$ and the maximum moves to the one of the surfaces of the slab. This can be interpreted as projection of the superconductivity on this slab surface}\label{fig1-MScIII}
\end{figure}

\begin{figure}[!ht]
\epsfxsize=\linewidth
\epsfbox{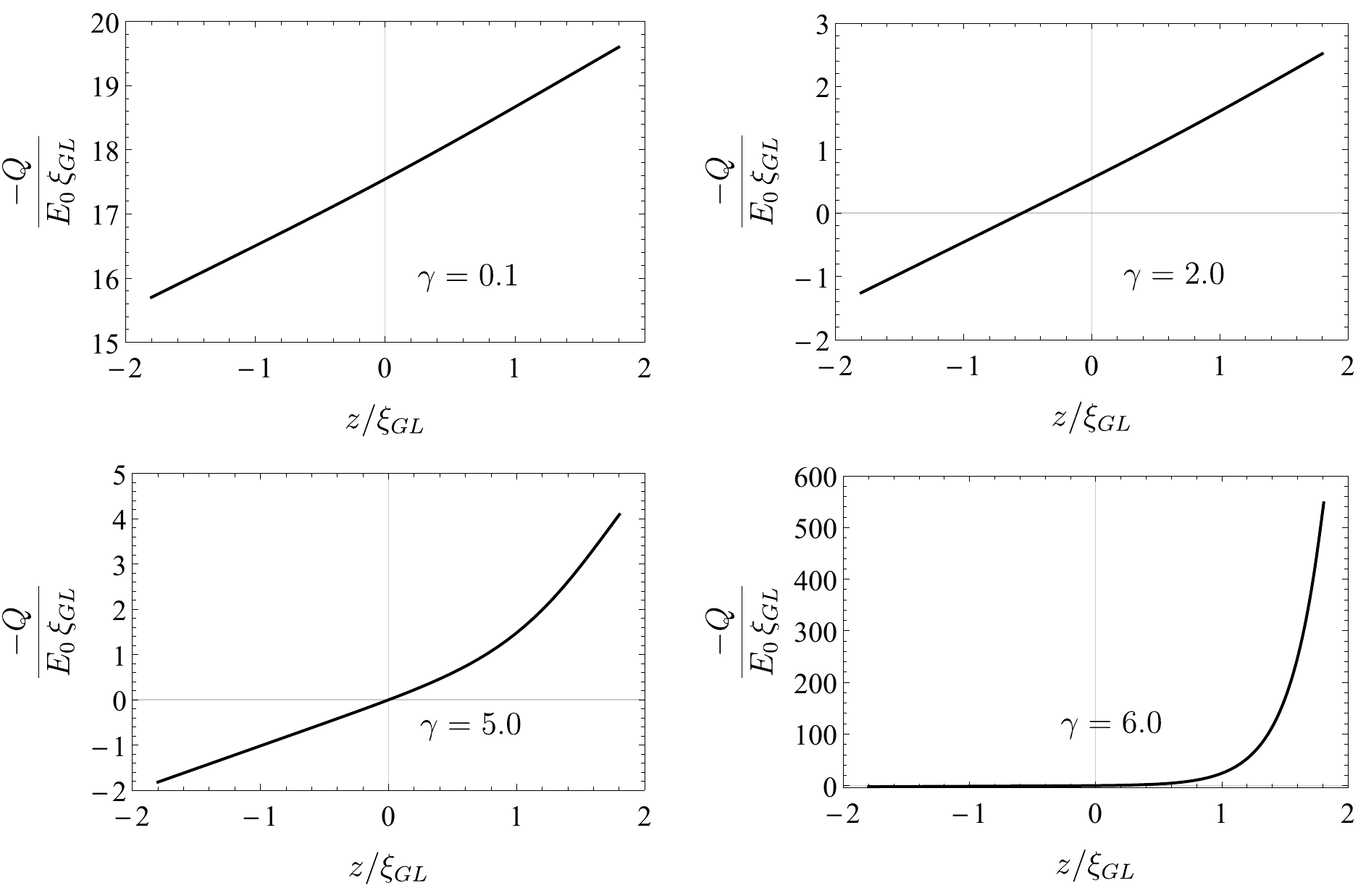} \caption{\,\,The dimensionless function $f_1=-\frac {Q}{E_0\xi_{GL}}$ of a dimensionless distance $\zeta=z/\xi_{GL}$, is plotted. Increasing $\gamma$ a characteristic distance $\zeta_0=z_0/\xi_{GL}$ occurs. Below this distance $f_1$ is negative, above the characteristic distance $f_1$ is positive  }\label{fig2-MScIII}
\end{figure}

\begin{figure}[!ht]
\epsfxsize=\linewidth
\epsfbox{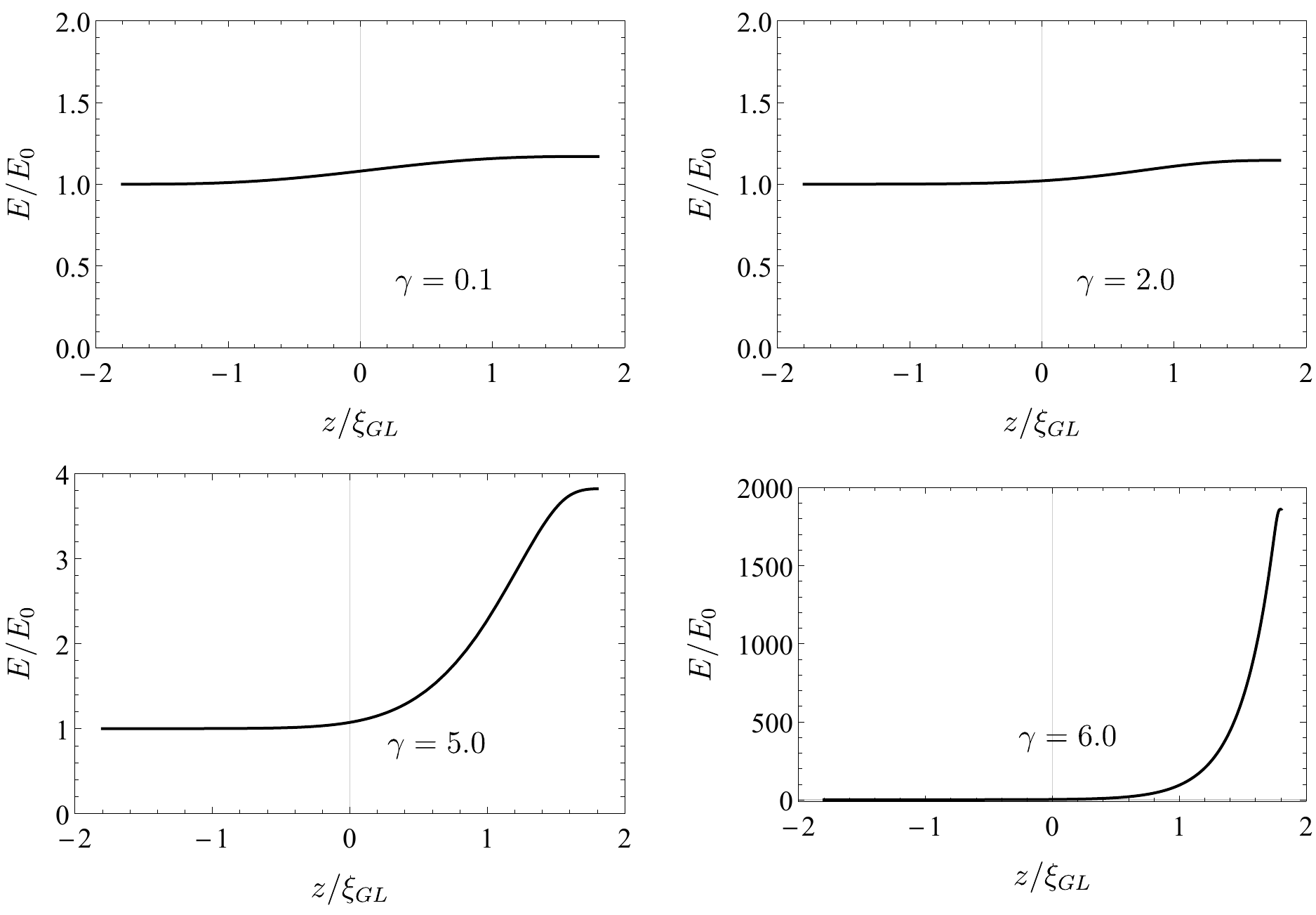} \caption{\,\,The dimensionless function $E/E_0$ of a dimensionless distance $\zeta=z/\xi_{GL}$, is plotted. The figures show that the strong coupling superconductors amplify the applied electric field}\label{fig3-MScIII}
\end{figure}

The dimensionless function $f_3=\rho/\rho_0$ of a dimensionless distance $\zeta=z/\xi_{GL}$ is depicted in figure (\ref{fig1-MScIII}) for different values of $\gamma$ proportional to applied electric field and $\kappa$ is fixed $\kappa = 2.5$. For slab geometry $\zeta$ runs the interval $-1.8\leq z/\xi_{GL}\leq 1.8$. The dimensionless function $f_2=E/E_0$ is depicted in figure (\ref{fig3-MScIII}) and
the dimensionless function $f_1=-\frac {Q}{E_0\xi_{GL}}$ is depicted in figure (\ref{fig2-MScIII}).

Increasing the applied electric field, keeping all parameters in the theory fixed, one increases the parameter $\gamma$. Figures (\ref{fig2-MScIII}) show that applying weak electric field (small $\gamma=0.1$) the function $f_1$ is positive so that $-f_3+\gamma f_1 f_3$ is positive, the Cooper pairs Bose condense $f_3>0$ and the onset of superconductivity starts while the temperature is higher then critical one $\alpha<0$. The function $f_3$ is small because the parameter $\gamma$ is small. There is no Bose condensation when $\gamma<0.1$.

Figures (\ref{fig2-MScIII}) show that increasing $\gamma$ a characteristic distance $\zeta_0=z_0/\xi_{GL}$ occurs. Below this distance $f_1$ is negative the density of Cooper pairs is very small even zero $f_3=0$. Above the characteristic distance, $f_1$ is positive the density function $f_3>0$ is positive. When $\gamma$ increases $\zeta_0$ increases too, the function $f_1$ abruptly increases near the surface $\zeta=1.8$ of the slab, which in turn leads to an abrupt increase of the Bose-condensate of Cooper pair near the surface.
This can be interpreted as a "projection" of the superconductivity on the surface.

In conclusion, we can say that studying the electrodynamics of s-wave superconductivity we arrived at the result that if the system has Cooper pairs, above the critical temperature, the applied electric field leads to the Bose condensation of fermion pairs and onset of superconductivity. Our finding offers new rout to check out the existence of Cooper pairs in a normal state.

The figures (\ref{fig3-MScIII}) show that the strong coupling superconductors amplify the applied electric field.

The earliest study of the electrodynamics of s-wave superconductors is attributed to London brothers \cite{London35}. The solutions of Maxwell-London equations showed that the electric and magnetic fields penetrate superconductor an equal distance of $\lambda_L$. The Londons' system of equations is relativistically covariant. In the papers \cite{Karchev17} and \cite{Vetsov17} we generalized this system considering relativistically covariant Ginzburg-Landau theory. The solutions of the system of equations, following from the relativistic Ginzburg-Landau theory, showed that the applied magnetic field suppresses the superconductivity, while the applied electric field supports it. It was reported that the magnetic and electric penetration depths are different. The discrepancy follows from the rough London approximation which does not account for the last two terms in Eq.\ref{EScCp15}  These terms are responsible for the different influence of the magnetic and electric fields on superconductivity.

The equations in relativistic theory \cite{Karchev17}, \cite{Vetsov17} are invariant under the discrete transformation $\textbf{B}\rightarrow-\textbf{B}$,
$\textbf{Q}\rightarrow-\textbf{Q}$ and independently under the transformation  $\textbf{E}\rightarrow-\textbf{E}$, $Q\rightarrow-Q$.
In contrast, the system of equations in nonrelativistic theory (\ref{EScCp20}-\ref{EScCp22}) is not invariant under the discrete transformation of electric field $\textbf{E}$ and gauge invariant field $Q$.
An important consequence is that when the applied electric field increases the maximum of $\rho/\rho_0$ as a function of the dimensionless distance $z/\xi_{GL}$ moves to one of the surfaces of the slab (Fig.\ref{fig1-MScIII}). We interpreted this as a "projection" of the superconductivity on this slab surface. This phenomenon is indicative of non-relativistic electrodynamics of superconductivity.

When the superconductor in the normal state is metal, near the critical temperature,
even in the superconducting phase, the density of normal quasiparticles is high and they screen the electric field. The density of normal quasiparticles decreases at low temperature and at very low temperatures is zero. The electric field penetrates the superconductor and important consequence is that increases the critical magnetic field. This result is experimentally testable.

\end{document}